# The scintillation of liquid argon

T. Heindl[1], T. Dandl[1], M. Hofmann[2], R. Krücken[1], L. Oberauer[2], W. Potzel[2], J. Wieser[3], and A. Ulrich[1]

[1]Physik Department E12, Technische Universität München, James-Franck-Str. 1, 85748 Garching, Germany
[2]Physik Department E15, Technische Universität München, James-Franck-Str. 1, 85748 Garching, Germany
[3]excitech GmbH, Branterei 33, 26419 Schortens, Germany

**Abstract.** A spectroscopic study of liquid argon from the vacuum ultraviolet at 110 nm to 1000 nm is presented. Excitation was performed using continuous and pulsed 12 keV electron beams. The emission is dominated by the analogue of the so called 2$^{nd}$ excimer continuum. Various additional emission features were found. The time structure of the light emission has been measured for a set of well defined wavelength positions. The results help to interpret literature data in the context of liquid rare gas detectors in which the wavelength information is lost due to the use of wavelength shifters.

Liquid rare gas detectors (LRGD) are widely used as particle and γ-radiation detectors [1, 2]. Both the scintillation light emitted in the vacuum ultraviolet (VUV) and the charge produced by the primary particle is used to detect and identify the primary particle. Liquid rare gases (LRG) are presently intensely studied and practically used as a material for detecting weakly interacting particles such as neutrinos or candidates for dark matter and to search for neutrinoless double beta decay. Such experiments require the installation of large detectors in underground laboratories with ultra high sensitivity and the capability to discriminate between the various particles interacting with the detector for efficient background reduction. Some detectors use the scintillation effect alone, caused by the primary particle in the liquid for particle discrimination [3,4], while others are built as 2 phase time projection chambers (TPC), in which the charge is also collected [5-7]. Wavelength shifters normally convert the VUV scintillation photons into ultraviolet (UV) or visible light where comparatively inexpensive photomultipliers can be used for light detection. Thereby, it is possible to record the time structure of the light emitted by the scintillating material when fast wavelength shifters are used. However, all the spectral information which may be very important for particle identification is lost with this technique. The routine use of wavelength shifters has led to the situation that, to the best of our knowledge, the spectral characteristic of the LRGs has not been studied for a very long time. Essentially all recent publications on this subject refer to work performed by J. Jortner and co-workers in the 1960ies and 1970ies [8,9]. However, the spectral range shown in that publication is limited to 115 − 180 nm. It is the purpose of this letter to report on the time structure of the scintillation light in liquid argon recorded over a broad wavelength region with good wavelength resolution investigating different emission features of the spectra. Wavelength resolved measurements can also help to detect and identify impurities present in the LRG, which show emission features at specific wavelengths. Adding admixtures to the LRG scintillator, deliberately and observing their time and wavelength resolved emission may help in the future to improve particle discrimination capabilities of liquid argon detectors. Wavelength resolved studies are therefore also important for studying this aspect.

The light intensity produced by laboratory radioactive sources is normally too weak for obtaining good quality optical spectra. We have overcome this problem with a table-top technique [10] in which a 12 keV electron beam is sent into liquid argon.

It is well known that the light emission from dense, particle beam excited rare gases is dominated by the so called "second excimer continuum" which is due to the radiative decay of the lowest lying molecular states $^1\Sigma$ and $^3\Sigma$ in the gas phase [11]. The analogue emission from the self trapped excitons is called the "M-band" in the solid [12]. Jortner had shown that this feature is also found in the liquid [8] and the time structure observed in the present LRGDs is always interpreted in terms of the singlet and triplet lifetimes tabulated as 4.2 ns and 3.2 µs, respectively for argon in the gas phase [11]. Being aware of the fact that the terminology may be inadequate for the liquid phase we will nevertheless refer to this emission as the "second continuum" in this paper and also adapt the terminology of the gas phase for other emission features.

This investigation was motivated by previous studies by some of the current authors of the light emission from particle beam excited gases, in particular the so called "third continuum" of the rare gas excimer emission [13]. For argon in the gas phase it can be much stronger than the 2$^{nd}$ continuum in the early phase following pulsed heavy ion beam excitation. This 3$^{rd}$ continuum also appears in solid argon [14]. In neon it is found in the gas phase with heavy ion excitation [15] but absent with electron beam excitation (see Fig. 3 in [16]). Therefore we were interested whether a corresponding feature might appear in the LRGs as well, in particular in argon and xenon which are used as detector material. There it might influence the interpretation of the time structure observed with wavelength integrated detection techniques such as wavelength shifters.

In the experiments described here, liquid argon was excited by a 12 keV, 1.5 µA electron beam sent through a 300 nm thick silicon nitride membrane into the liquid. Light exiting the target cell through a $MgF_2$ window was collected with an ellipsoidal mirror, analyzed with a f=30 cm VUV monochromator (McPherson 218, Grating: 1200 mm$^{-1}$, Blaze 150 nm) and recorded by photon counting using a VUV photomultiplier with $MgF_2$ window and S20 cathode. The wavelength dependence of the relative response of the detection system was calibrated against a deuterium lamp with a calibrated emission spectrum. In the present setup the electron beam could be pulsed with a pulse width of 200 ns and a repetition rate of 10 kHz. Time spectra were recorded using a standard time to amplitude conversion technique. The liquid argon cell was connected via a heat exchanger to a gas system at room temperature with a gas purifier [SAES Getters, MonoTorr® Phase II, PS4-MT3] and a metal-bellows pump which continuously circulated the gas through the system. Details of the setup will be described in a separate publication.

A spectrum between 115 and 320 nm of the emission from liquid argon recorded with continuous electron beam excitation in comparison with a spectrum recorded with the same apparatus for 0.3 bar argon in the gas phase is shown in Fig. 1. We use this comparison for a tentative assignment of the light emission features found in liquid argon. The spectrum in the liquid is dominated by an emission analogue to the 2$^{nd}$ continuum band in gas phase. Its intensity peaks at 126.8 nm and its full width at half maximum is 7.8 nm. The weak structure around 155 - 160 nm can be addressed as the analogue of the so called "classical left turning point" emission [17]. The continuum beyond 170 nm is known as the 3$^{rd}$ excimer continuum [13] and extends up to ≈300 nm. Liquid argon also shows some emission in this region. The overall light output, however, is weak representing less than 0.2% of the total light output between 100 and 320 nm. Since the trend in the gas phase is such that the 3$^{rd}$ continuum shifts to longer wavelengths in denser gas one could argue that the peak at 270 nm in the liquid phase has a similar origin as the 3$^{rd}$ continuum described in Ref. [13].

Using a sensitive spectrometer (OCEAN Optics HR65000) we also recorded light in the spectral region between 250 and 1000 nm. The spectrum is shown in Fig. 2. Calibration of the spectrometer was done using a calibrated halogen lamp. We found weak, but various emission features in this spectral region with presently unknown origin. The peak at 557 nm is caused by the emission of an oxygen impurity [12]. We can observe a broad continuum underlying the whole spectrum so that the liquid argon emission appears as a whitish glow to the human eye. The emission at 970 nm close to the cut-off wavelength of our spectrometer indicates that liquid argon might have some rather strong emission features in the near infrared.

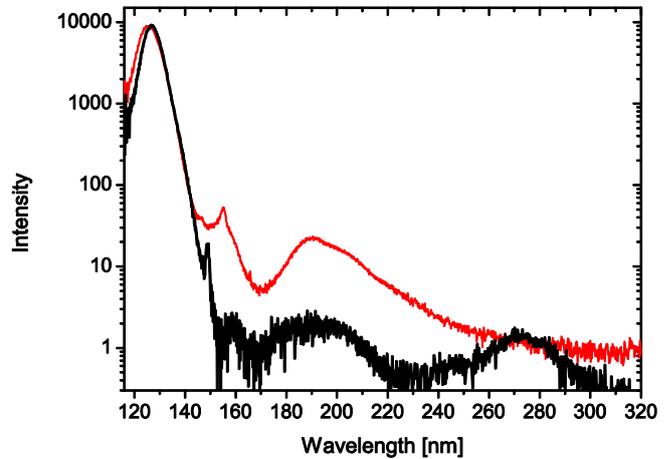

*Fig. 1.* VUV/UV emission spectrum of liquid argon (85 K, thick line) in comparison with gaseous argon (295K, 300 mbar, thin line). The liquid argon spectrum is dominated by an emission feature (126.8 nm) analogue to the 2$^{nd}$ excimer continuum in the gas phase. Weak emission features in the wavelength range from 145 to 300 nm can be observed. The peak at 149.1 nm in liquid argon is caused by a xenon impurity in our gas-system (see text). The structure at 155 nm in the gas phase which is called "classical Left Turning Point" LTP in the literature has only a very weak analogue in the liquid phase. The structure at longer wavelengths up to 320 nm is addressed as the 3$^{rd}$ continuum emission in the gas phase. Note, that the sensitivity of the detection system has been calibrated in the region between 115 and 230 nm.

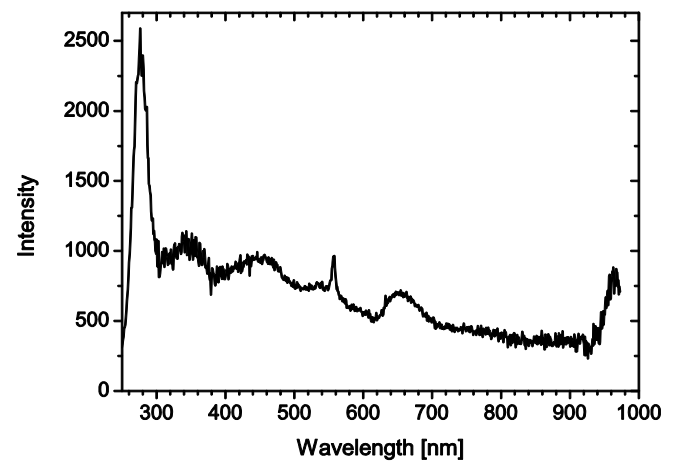

*Fig. 2.* Emission spectrum of liquid argon (85 K) from 250 to 970 nm. The sensitivity of the spectrometer was calibrated with a halogen lamp. The emission structure at 270 nm (also shown in Fig. 1) is the long wavelength part of the analogue of the 3$^{rd}$ Ar continuum. The peak at 557 nm is caused by the emission from a residual oxygen impurity [12].

A time spectrum emitted from the 2nd continuum in liquid argon is shown in Fig. 3. It was recorded at the peak wavelength of 126.8 nm and various nearby positions with a spectrometer slit width of 200 μm which corresponds to a bandwidth of 0.5 nm. Data at different wavelengths within the 2nd continuum emission showed no variation with wavelength. A fast ($\tau$ < 6.2 ns) and a slow decay time are observed following the end of a 200 ns excitation pulse. We measure a slow decay time of (1300 ± 60) ns. This value is comparable with literature data, where decay times of (1590 ± 100) ns [18] and 1100 ns [19] are measured. For a list of reported values see TABLE II in Ref. [20] and references therein. The time resolution of the experimental system will have to be improved to measure the fast decay time with good resolution. The time structure of the analogue of the 3rd continuum was also measured at a wavelength of 195 and 270 nm and it was found that the emission follows the excitation promptly within the time resolution of approximately 6 ns of the present setup. Note that the 3rd continuum emission is also fast in the gas phase [13].

the spectrum. With the xenon admixture further increased the argon continuum is modified in a way, that it extends to the longer wavelength side. Furthermore another broad feature appears in the spectra around 175 nm which is the analogue of the 2nd continuum emission of the xenon excimer while the 3rd continuum emission of argon disappears. For higher xenon concentration the 2nd continuum emission of the xenon excimer seems to become the dominant emission in the spectrum [9,21,22].

A time spectrum recorded on the xenon resonance line at 149.1 nm is shown in Fig. 4. The emission time is very long with an intensity increase over the first 2 μs. This increase indicates that there must be an intermediate excitation step involved. In the gas phase similar time dependences are e.g. found for emission following recombination with free electrons which have to cool after the excitation pulse thereby increasing the recombination rate [23]. Note that the slow decay time of the 2nd continuum is unaffected by the impurity, at least at a low level of the xenon impurity.

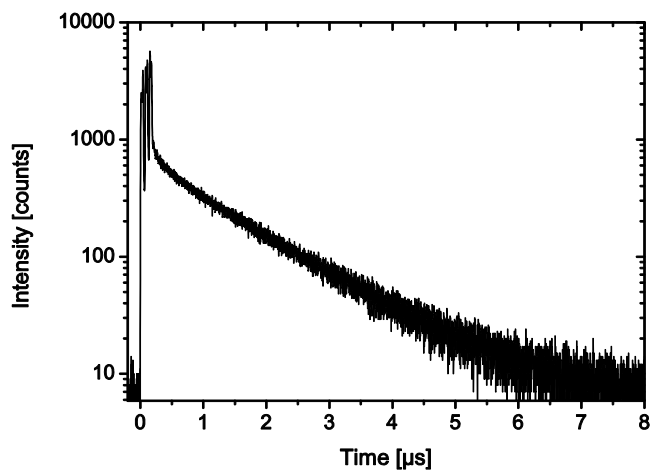

**Fig. 3.** Time spectrum emitted from the 2nd continuum of liquid argon following excitation with an electron beam pulse of 200 ns duration.

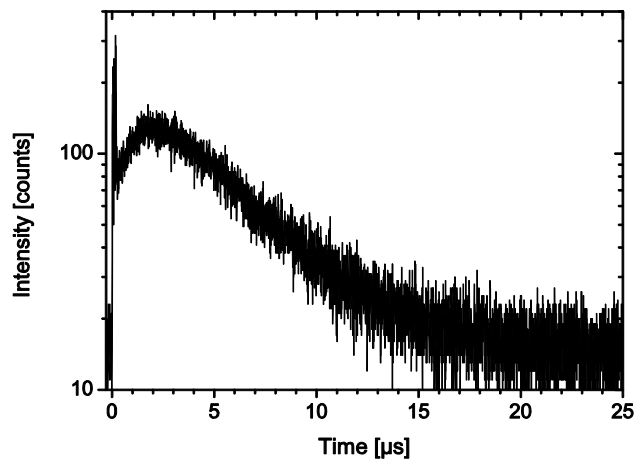

**Fig. 4.** Time spectrum emitted from a xenon impurity in liquid argon at a wavelength of 149.1 nm. The electron beam excitation pulse had a duration of 200 ns.

As in the gas phase the VUV emission spectrum of argon is extremely sensitive to impurities also in the liquid phase. Heavier rare gases such as xenon and typical impurities like oxygen, nitrogen and carbon may play an important role. The latter were removed in our system by a rare gas purifier prior to filling as well as during operation by evaporating and re-condensing the gas in a closed cycle. Removing a xenon impurity is difficult because it is not removed by a rare gas purifier. After month-long repetitive flushing and operation of the system, which had been used with xenon before, the purifier was replaced to get rid of xenon. But even after that traces of xenon still contributed to the emission spectrum, as can be seen in Fig. 1. We observe that the xenon resonance line (at 147.0 nm in the gas phase) is shifted to 149.1 nm in the liquid [9]. With an increasing impurity of xenon in the liquid argon, the emission structure at 149.1 nm increases without any observable changes in the rest of

Evaporation of argon due to the power deposited by the electron beam can "contaminate" the liquid with gaseous argon. Therefore, the electron beam current was always limited to 1.5 μA (cw operation) in our experiments. Higher beam currents led to evaporation of the gas which can be observed by a strong appearance of the 4p-4s ArI lines in the red and near infrared. The absence of the ArI lines was also used in the experiments with pulsed excitation as an indication for undisturbed conditions.

For practical application it is interesting to study how the emission spectrum changes when the gas purification is not used in a nominally very clean gas system as the one described above. The result for our specific system after filling and condensing argon directly from the bottle (Ar 4.8, Linde AG) is shown in Fig. 5. The strong emission with peak intensity at 197 nm and 16.2 nm width may tentatively be attributed to an oxygen impurity based on a discussion published in Refs. [12,24]. The spectrum is plotted in Fig. 5 with and without sensitivity correction of the detection system to demonstrate that the signal may be very strong in all detector-systems where the sensitivity drops towards shorter wavelengths as it is normally the case. The time structure for the 197 nm impurity is qualitatively similar to the one observed for the xenon impurity shown in Fig. 3 and extends over roughly 15 µs. Here, however, the decay time of the argon 2$^{nd}$ continuum light is strongly affected by the impurity and reduced from 1300 to 342 ns for the impurity level shown in Fig. 5.

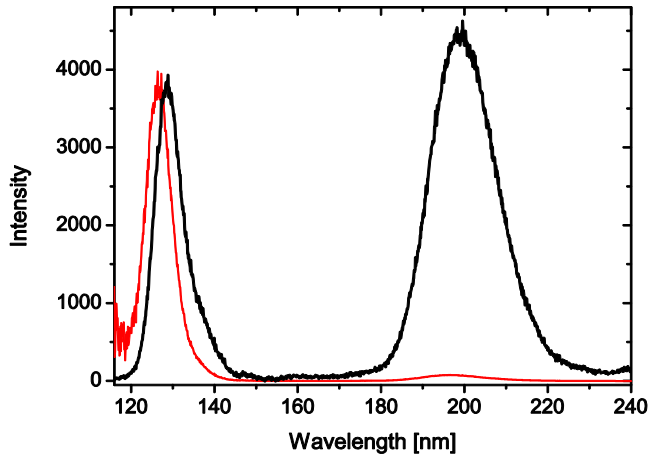

**Fig. 5.** *Emission spectrum of unpurified liquid argon (85K) with (thin line) and without (thick line) sensitivity correction of our setup. The broad emission structure around 197 nm may be attributed to an oxygen impurity [12, 24].*

In summary, we have performed a study providing an overview over the scintillation of electron beam excited liquid argon. The results show, that for electron beam excitation an emission structure analogue to the 2$^{nd}$ excimer continuum in the gas phase indeed dominates the spectra. In that case wavelength shifters can be used in pure liquid argon detectors, without loosing significant information. If this is also true for other projectiles, will be a subject of further investigations. Experiments to study the influence of the mass and charge of the projectiles are in preparation. Various heavy ion beams from the Munich Tandem van de Graaff accelerator will be used for that purpose. In a recent publication [4] it was shown, that the particle discrimination capability of wavelength shifted liquid argon detectors can be improved by adding small amounts of xenon. Since the xenon admixture modifies the spectral emission, systematic studies with variable xenon admixture in our setup will help to boost those capabilities further by making use of the wavelength information. Similar studies will also be performed with better time resolution, and for other rare gases, mainly xenon.

## Acknowledgement

We thank Dr. Teresa Marrodán for stimulating discussions of the subject and funding by the Maier Leibnitz Laboratory Munich (MLL), the German Ministry of Education and Research (BMBF) No. 13N9528, and the Deutsche Forschungsgemeinschaft DFG (Transregio 27: Neutrinos and Beyond).